\documentclass[%
reprint,
superscriptaddress,
 amsmath,amssymb,
 aps,
 prx,
floatfix,
]{revtex4-1}
\usepackage[utf8]{inputenc}
\usepackage{graphicx,color,soul} 
\usepackage{amsmath}
\usepackage{hyperref}
\usepackage{upgreek}

\begin{document}
\title{A Polariton-Stabilized Spin Clock}




\author{Matthew E. Trusheim} 
\affiliation{U.S. Army Research Laboratory, Sensors and Electron Devices Directorate, Adelphi, Maryland 20783, USA} 
\affiliation{Department of Electrical Engineering and Computer Science, Massachusetts Institute of Technology, 77 Massachusetts Avenue, Cambridge, Massachusetts 02139, USA}

\author{Kurt Jacobs}
\affiliation{U.S. Army Research Laboratory, Sensors and Electron Devices Directorate, Adelphi, Maryland 20783, USA}
\affiliation{Department of Physics, University of Massachusetts at Boston, Boston, MA 02125, USA}

\author{Jonathan E. Hoffman} 
\affiliation{U.S. Army Research Laboratory, Sensors and Electron Devices Directorate, Adelphi, Maryland 20783, USA}

\author{Donald P. Fahey} 
\affiliation{U.S. Army Research Laboratory, Sensors and Electron Devices Directorate, Adelphi, Maryland 20783, USA}

\author{Danielle A. Braje} 
\affiliation{MIT Lincoln Laboratory, Lexington, MA 02420, USA}

\author{Dirk Englund} 
\affiliation{Department of Electrical Engineering and Computer Science, Massachusetts Institute of Technology, 77 Massachusetts Avenue, Cambridge, Massachusetts 02139, USA}

\date{July 2020}

\begin{abstract} 
Atom-like quantum systems in solids have been proposed as a compact alternative for atomic clocks, but realizing the potential of solid-state technology will requires an architecture design which overcomes traditional limitations such as magnetic and temperature-induced systematics. Here, we propose a solution to this problem: a `solid-state spin clock' that hybridizes a microwave resonator with a magnetic-field-insensitive spin transition within the ground state of the diamond nitrogen-vacancy center. Detailed numerical and analytical modeling of this `polariton-stabilized' spin clock (PSSC) indicates a potential fractional frequency instability below $10^{-13}$ at 1 second measurement time, assuming  present-day experimental parameters. This stability would represent a significant improvement over the state-of-the-art in miniaturized atomic vapor clocks. 

\end{abstract}

\date{\today}

\maketitle
\section{Introduction}

A longstanding goal in measurement science and technology is the miniaturization of precision oscillators based on atomic states. Laboratory-scale clocks based on atomic vapors can reach state-of-the-art fractional frequency deviations reaching $10^{-15}$\cite{Martin2018-es,Micalizio2012-xq}, while scaled-down commercial atomic vapor clocks achieve a fractional frequency stability $\Delta \nu/\nu \sim  10^{-10}$ at one second of measurement time in a volume of order 10 cm$^3$, representing orders-of-magnitude improvements in size, mass, and power. Still further  reduction may be possible by transitioning to chip-integrated, atom-like solid-state systems operating at room temperature.
 
 To this end, recent work has investigated frequency references based on the diamond nitrogen vacancy center~\cite{Hodges2013-qx, Breeze2018-wm}, whose spin-1 electronic ground states ($|m_s=\pm 1,0\rangle$) have coherence times at exceeding milliseconds at root temperature\cite{Stanwix2010-xl,Bradley2019-vc}, an $m_s=0 \rightarrow \pm 1$ zero-field splitting near 2.87 GHz, near-unity optical polarization under ambient conditions\cite{Doherty2013-ay}, and high-fidelity readout by fluorescence~\cite{Clevenson2015-xk,Barry2020-zk} or microwave absorption~\cite{Eisenach2020-cu,Ebel2020-ih}. Moreover, the spin-1 system allows for magnetic-field-insensitive coherence between the $|+\rangle = |-1\rangle+|1\rangle$ and $|0\rangle$ levels, analogous to an atomic clock transition\cite{Toyli2013-nv,Brunner2013-em}. Hodges~\textit{et al}.\ previously proposed the use of this transition in an ensemble of $\sim10^8$ NV centers to produce an clock with shot-noise limited fractional frequency stability of $\delta\nu/\nu=10^{-12}$ at 1 second --- \textit{if} temperature could be sufficiently stabilized. Indeed, converting the performance of demonstrated NV magnetometers with sensitivities $\sim 10^{-12}$ T Hz$^{-1/2}$~\cite{Eisenach2020-cu,Wolf2015-ze,Chatzidrosos2017-od,Schloss2018-ce} to a fractional frequency deviation based on the NV gyromagnetic ratio (28 GHz/T at 3 GHz) implies $\delta\nu/\nu = 10^{-11}/\sqrt{Hz}$. Realistically achieving this stability for long integration times, however, would require temperature control at the $\upmu$K level or technically demanding differential-clock tracking schemes~\cite{Hodges2013-qx} that have yet to be demonstrated. At a realistic thermal instability at the mK level, the absolute fractional frequency precision of an uncompensated NV-diamond clock is near $\delta \nu/\nu \sim 10^{-7}$ and thus not competitive.  

Here we address this crucial problem by reducing the oscillator temperature dependence in a hybrid timekeeping device: a polaritonic system consisting of coupled NV spins and microwave cavity photons that produces an avoided crossing, removing the first-order thermal response and enabling better-than-mHz stability for realistic spin-photon coupling strengths. Through analytical and numerical modeling of the polariton response and spin-cavity input-output relations, we show that the thermal and magnetic limits to fractional frequency stability can reach below $10^{-13}$ with realistically achievable stabilization. This potential improvement of over five orders of magnitude could enable a new class of solid-state, chip-integrated spin-based clocks with applications from positioning, navigation, and timing to instrumentation and telecommunications.

\section{Scheme} 


\begin{figure}
\centering
    \includegraphics[width=3.4in]{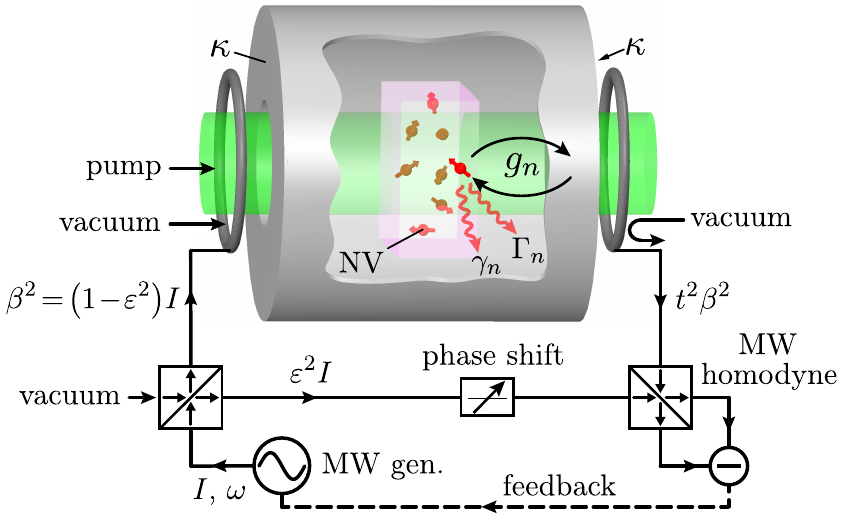}
	\caption{A polaritonic-stabilized spin clock. The reference probe field at frequency $\omega$ with intensity $I$ is is split into a local oscillator and probe. The latter is sent to the resonant system, which consists of NV centers in diamond with population relaxation rate $\gamma$ coupled to a microwave cavity at rate $g$. The cavity mirrors have effective loss rates $\sqrt{\kappa}$, and the transmitted field is detected as the signal via homodyne measurement. \label{fig1}}
\end{figure}

Figure 1 illustrates the proposed clock, consisting of an ensemble of $N$ diamond NV centers coupled to a microwave cavity\cite{Eisenach2020-cu,Ebel2020-ih}. Each color center $n$ has two magnetic resonance transitions $\omega_{n,\pm}$ corresponding to its $m_s = |0\rangle$ to $m_s = |\pm\rangle$ transitions within the NV ground state spin triplet, with spin-cavity coupling rates $g_{n,\pm}$ and dephasing rates $\Gamma_{n,\pm}$. The cavity resonance at frequency $\omega_c$ has loss rate $\kappa$ and is probed continuously by a microwave field at frequency $\omega$. The NV is continuously optically polarized~\cite{Robledo2011-sa}, which we model as damping at a rate of $\gamma_{n,\pm}$ that keeps the NV ensemble close to the $|0\rangle$ state. For a weak, continuous probe field $\beta$, we use the input-output formalism~\cite{Jacobs14,Wiseman2010-hl} 
to derive a closed-form expression for the transmitted field $t\beta$~\footnote{Details of the input-output analysis for the ensemble of three-level systems in the microwave cavity are given in the supplemental information}, in which   
\begin{align} 
   t & =  \frac{\kappa}{ \kappa + \kappa_{{l}} + i \Delta  + C}, 
\end{align}
with $\Delta = \omega_c - \omega$, $\kappa_{{l}}$ is the internal cavity loss rate, $C$ sums over all the spin transitions of all $N$ NV centers:
\begin{align} 
C & = \sum_{n,j=\pm}\frac{g_{n,j}^2}{(\Gamma_{n,j}+\gamma_{n,j})/2 + i\Delta_{n,j}}, 
\end{align}
and $\Delta_{n,j} = \omega_{n,j}-\omega$. 


Homodyne detection is implemented by splitting the initial microwave source with power $I$ into two beams, the probe and a reference, and then mixing the reference with the cavity output as depicted in Fig.\ref{fig1}.  This scheme allows for measurement of any quadrature of the cavity output field depending on the phase shift applied to the reference\footnote{See supplemental information}. Fig.\ 2 plots the the optimally-sensitive phase quadrature $\operatorname{Im}[t]$ for the coupled spin-cavity system under the realistic parameters summarized in Table 1. We assume equal dephasing rates for each spin sublevel $\Gamma_{n,\pm} = \Gamma$, equal optical polarization rates $\gamma_{n,\pm} = \gamma$ and equal coupling $g_{n,\pm} = g_0$, achieved via a properly-polarized cavity mode along the crystalline [100] direction. Figure 2a assumes an external magnetic field oriented along the [100] crystalline axis, producing a Zeeman splitting of $\Delta_\pm$ = 10 MHz between the $m_s = \pm1$ states. The transmission shows a clear normal-mode splitting as the cavity matches the spin resonance frequencies $\omega_+$ and $\omega_-$. These are the energies of the spin-cavity polariton modes, which were experimentally observed recently for an NV ensemble coupled to a dielectric microresonator cavity~\cite{Eisenach2020-cu}. 

\begin{figure}
	\includegraphics[width=0.5\textwidth]{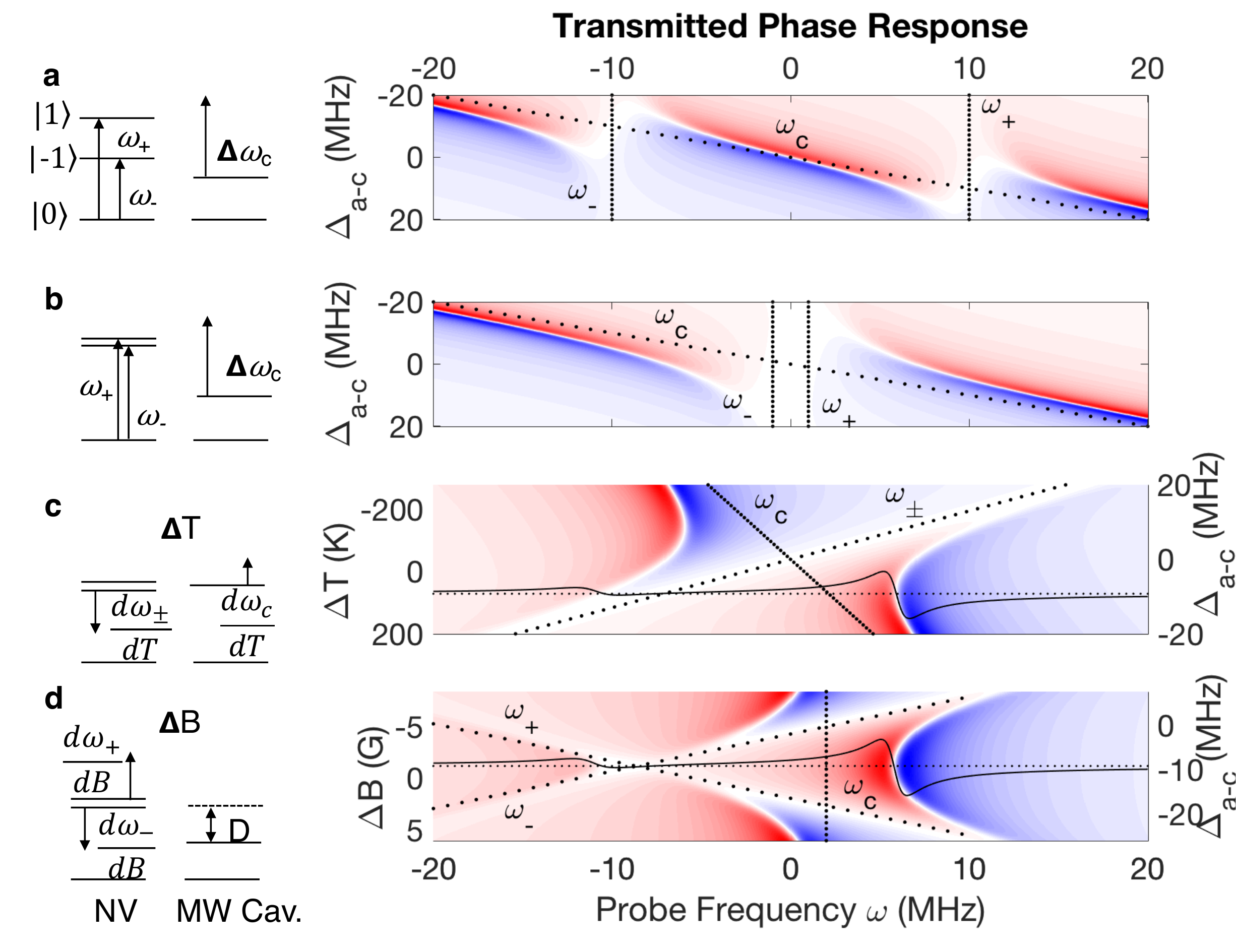}
	\caption{Transmission spectra. (a) In the presence of a moderate magnetic field, the NV transitions $\omega_i = \pm10$ MHz are split, resulting in three resonant polariton branches as the NV-cavity detuning is swept. (b) At low fields, the NV transitions $\omega_i = \pm1$ MHz are nearly degenerate and two polariton branches are visible. (c) Temperature shifts the NV and cavity in opposite directions, with a relative magnitude of $R=0.3$ shown here. The polariton branches have a vanishing derivative with respect to temperature at a detuning $D = 9.25$ MHz, while maintaining a sharp change in phase with probe frequency indicated by the black curve. (d) Magnetic-field dependent spectra taken at  the thermally-insensitive operating point. The zero-field spin-cavity detuning $D = 9.25$ MHz, and $\omega_c = 2$ MHz for direct comparison with (c) (black curve). At zero field there is a vanishing change in polariton frequency with magnetic field.  All figures use $\kappa = 500$ kHz, $\gamma = 3$ MHz, and $g = 5$ MHz. White corresponds to zero transmission.}
\end{figure}

The solid white contours in Fig.\ 2a closely follow the eigenfrequencies of the cavity-spin polariton energies $\lambda_j$ in the limit of low atomic damping. The corresponding eigenstates are the dressed states of a three-level spin system coupled by a classical driving field, which shows characteristic avoided crossings as the bare spin and cavity are nearly resonant. These avoided crossings merge into one as the magnetic field vanishes (see Fig.\ 2b). If the probe oscillator frequency deviates from the polariton resonance, the non-zero output phase is used as a feedback signal to correct the oscillator and provide long-term stability. The modified cavity transmission spectrum due to atom-cavity hybridization will allow cancellation of the primary source of temperature dependence of the polariton frequency to first order.
 
To illustrate the temperature insensitivity of the PSSC, we model the thermal response of the NV and cavity at room-temperature as linear with  coefficients $\frac{d\omega_a}{dT} = 77$ kHz/K\cite{Acosta2010-ar,Chen2011-td} and $\frac{d\omega_c}{dT}= R\frac{d\omega_a}{dT}$,  where $R$ defines their ratio. Fig.\ 2c depicts $\operatorname{Im}[t]$ for a room-temperature detuning $\Delta T$ between -200 to 200 K for a typical ceramic microwave resonator characterized by $R\sim 0.3$. The transmission shows a temperature-insensitive operating point at an atomic-cavity detuning $\Delta_{a-c} = 9.1$ MHz where the change in polariton frequency with temperature vanishes. By locking $\omega$ to this operating point, it is thus possible to cancel the temperature dependence to first order. 

This operating point retains the magnetic-field insensitivity of the $|+\rangle = |-1\rangle+|1\rangle$ and $|0\rangle$ transition. Fig.\ 2d, depicts $\operatorname{Im}[t]$ as a function of magnetic field and shows a vanishing change at an operating point of 0 G. This first-order frequency stability against temperature and magnetic field shifts is the key result underpinning the polariton-stabilized spin clock. We next consider how well this clock works under realistic assumptions of stabilizing against environment fluctuations.

\section{Performance} 

The critical figure of merit is the change of polariton frequency with temperature, $d\nu /dT$. We calculate this value from the analytical model of the cavity-spin system (see supplemental information), which in the limit of small spin Zeeman shift ($\omega_1 = \omega_2$), gives a simplified solution for the polariton energies 
\begin{align}
 \nu_{\pm} & = \frac{(\omega_c+\omega_a)}{2} \pm \sqrt{(\omega_c - \omega_a)^2/4 +  (g_+^2+g_-^2) }.
\end{align}
  
   Assuming equal coupling coefficients $g_+=g_- = g$, the polariton thermal dependence is
   
\begin{align}
 2\frac{d\nu_{\pm}}{dT} & = \frac{d\omega_a}{dT} + \frac{d\omega_c}{dT} \pm \frac{ (\frac{d\omega_a}{dT} - \frac{d\omega_c}{dT})(\omega_c - \omega_a)}{\sqrt{(\omega_c - \omega_a)^2 +  8g^2}}.
\end{align}
    
This yields a large reduction in $d\nu_{\pm}/dT$ in the regime where the coupling is greater than the thermal shift, as shown in Figure 3a. For large temperature shifts, the polariton thermal dependence approaches that of the uncoupled spin or cavity system. The thermal dependence $d\nu/d T$ approaches zero, however, at a particular spin-cavity detuning $\Delta_{a-c}=D_\pm$ as long as the thermal coefficients are opposite in sign, $R < 0$: 
    
\begin{align}
D_\pm & = \pm\sqrt{2}g\left[\sqrt{|R|}-\frac{1}{\sqrt{|R|}}\right].
\end{align}

Near this operating point, the polariton resonance frequency $\nu_\pm$ has only second-order dependence on temperature, resulting in a linear $\frac{d\nu_{\pm}}{dT}|_D\propto \frac{|R|^{3/2}}{g(|R|+1)^2}$. Figures 3b and 3c show the change in polariton frequency for a given temperature shift relative to the operating point, $\Delta\nu$, for various assumptions of the relative thermal responses $R$ and coupling rates $g$. In Fig. 3b, we take $R = 1$ and sweep $g$ from 0 to 5 MHz. A insensitive operating point can be see as $d\nu/dT = 0$  $D_+\implies\omega_c = \omega_a$ for all values of $g$ with the described quadratic energy dependence (linear derivative). Figure 3c shows the thermal response about the operating point $D_+$ for varying $R$ with a fixed $g =1$ MHz. Finally, Fig.\ 3d shows the limits of frequency accuracy as a function of coupling for different thermal and magnetic instabilities. A limit in the range of mHz is could be achievable for magnetic instability on the order of nT and thermal instability on the order of mK. 

\begin{figure}[h]
	\includegraphics[width=0.5\textwidth]{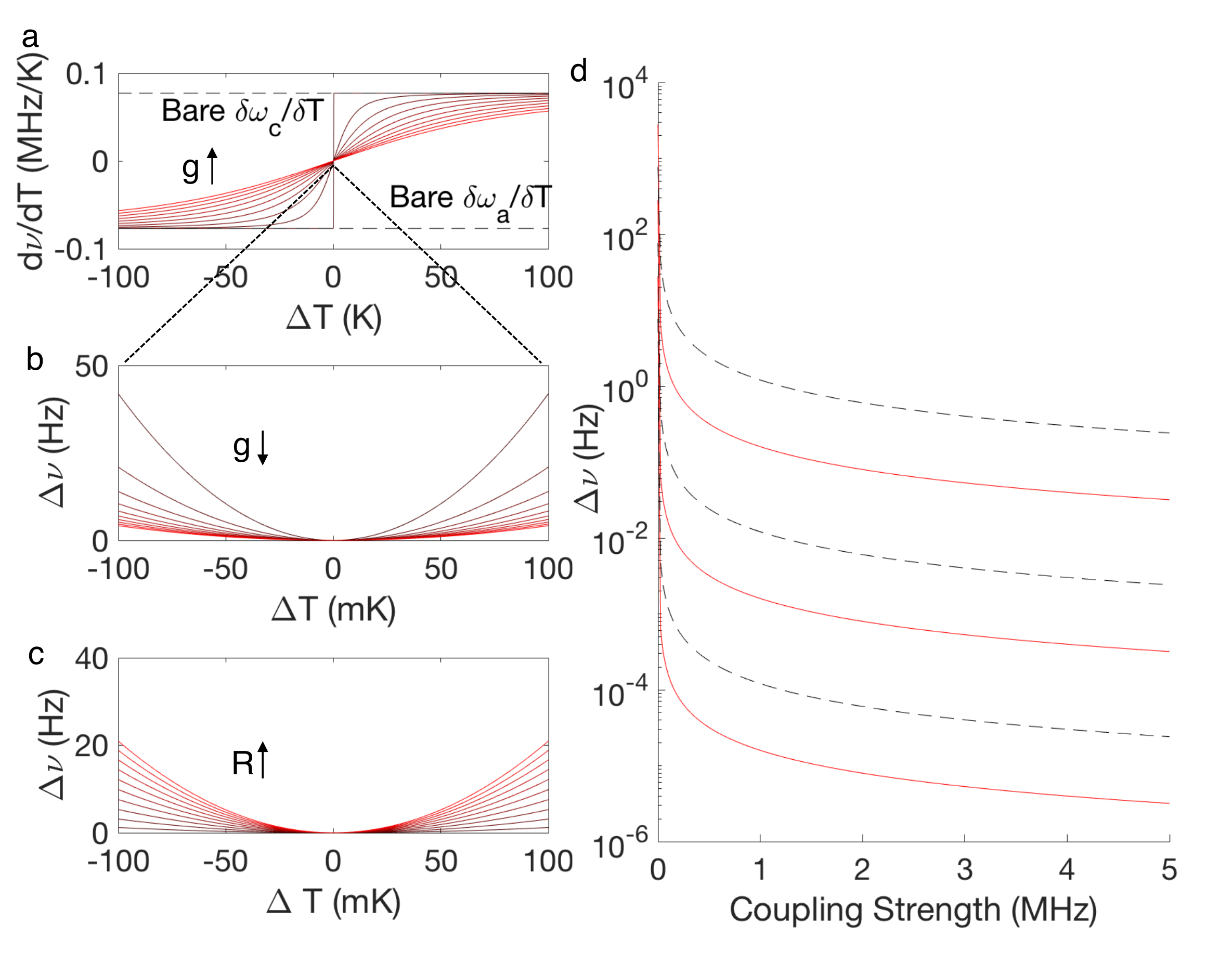}
	\caption{Thermal stability. (a) Thermal response of the polariton resonance frequency assuming $R=1$. Colored lines correspond to an increase in coupling strength from $g=0$ to $g=5$ MHz. Zero relative frequency corresponds to the NV zero-field splitting. (b) Polariton resonance frequencies as a function of temperature about the zero-response operating point. Lines indicate changes in coupling $g=0.5$ to $5$ MHz. (c) Polariton resonance frequencies as a function of temperature about the zero-response operating point, with $R$ varying from 0.1 to 1. (d) Polariton frequency error resulting from thermal and magnetic instabilities. Dashed lines: temperature instability descending logarithmically from 100 to 1 mK. Solid lines; magnetic field instability descending logarithmically from 100 to 1 nT.}
\end{figure}

A key figure of merit for clocks is the achievable fractional frequency deviation, $\delta \nu/\nu$. In addition to the oscillator stability, the fractional frequency deviation is also affected by the achievable measurement signal-to-noise ratio in a given measurement time. The signal measured in the polariton clock is the transmitted microwave field quadrature detected via homodyne readout, the noise for which is set by fundamental quantum processes. We perform a full noise analysis of the homodyne detection using input-output theory, which is detailed in the supplemental information. In the limit of low excitation and far detuned NVs, the precision of a frequency measurement achieved after an integration time $\tau$ is approximately
\begin{align}
    \delta\nu = \left(\frac{\kappa}{\sqrt{\tau I }} \right)     \sqrt{  1 + \frac{\xi^2}{2} }
\end{align}
in which $I$ is the power of the microwave source in photons per second, $\kappa$ is the cavity linewidth, and $\xi = \kappa_{{l}}/\kappa$ is the ratio of the internal loss rate of the cavity to its output rate. Finite detuning of the NVs, $D \lesssim \Gamma$, increases the error by a factor of order unity~\cite{Note1}. Improved performance is achieved by increasing the probe power and decreasing the cavity loss rate, but these parameters are limited by the assumption of low excitation of the NV centers given by  
\begin{align}
  \beta & \ll \frac{\sqrt{\kappa \gamma_{n,j} (\gamma_{n,j} + \Gamma_{n,j})}}{4 g_{n,j} |t|} , \;\; \forall n,j , 
\end{align} 
in which $|t|$ is of order unity. This limit is reached at around $\beta^2 \sim 10^{20}$ photons/sec for experimental parameters~\cite{Eisenach2020-cu}. Given the resulting measurement SNR, the polariton resonance can be determined to a fractional precision $\delta\nu/\nu$ after a measurement time $\tau$ of $\frac{\kappa}{\omega_a\cdot\mathrm{SNR}(\tau)} < 10^{-13}$ T Hz$^{-1/2}$. 

Environmental fluctuations, however, will limit the minimum achievable frequency deviation at long integration times. Figure 4a shows the minimum frequency deviation given several realistic levels of environmental instability  as a function of spin-cavity coupling. Combining these processes, Figure 4 plots the predicted fractional frequency deviation as a function of time, indicating that with an achievable temperature stability of 1 mK, the polaritonic clock  reaches the thermal stability limit below the $10^{-14}$ level after a few hundred seconds of averaging.

\begin{figure}
	\includegraphics[width=0.4\textwidth]{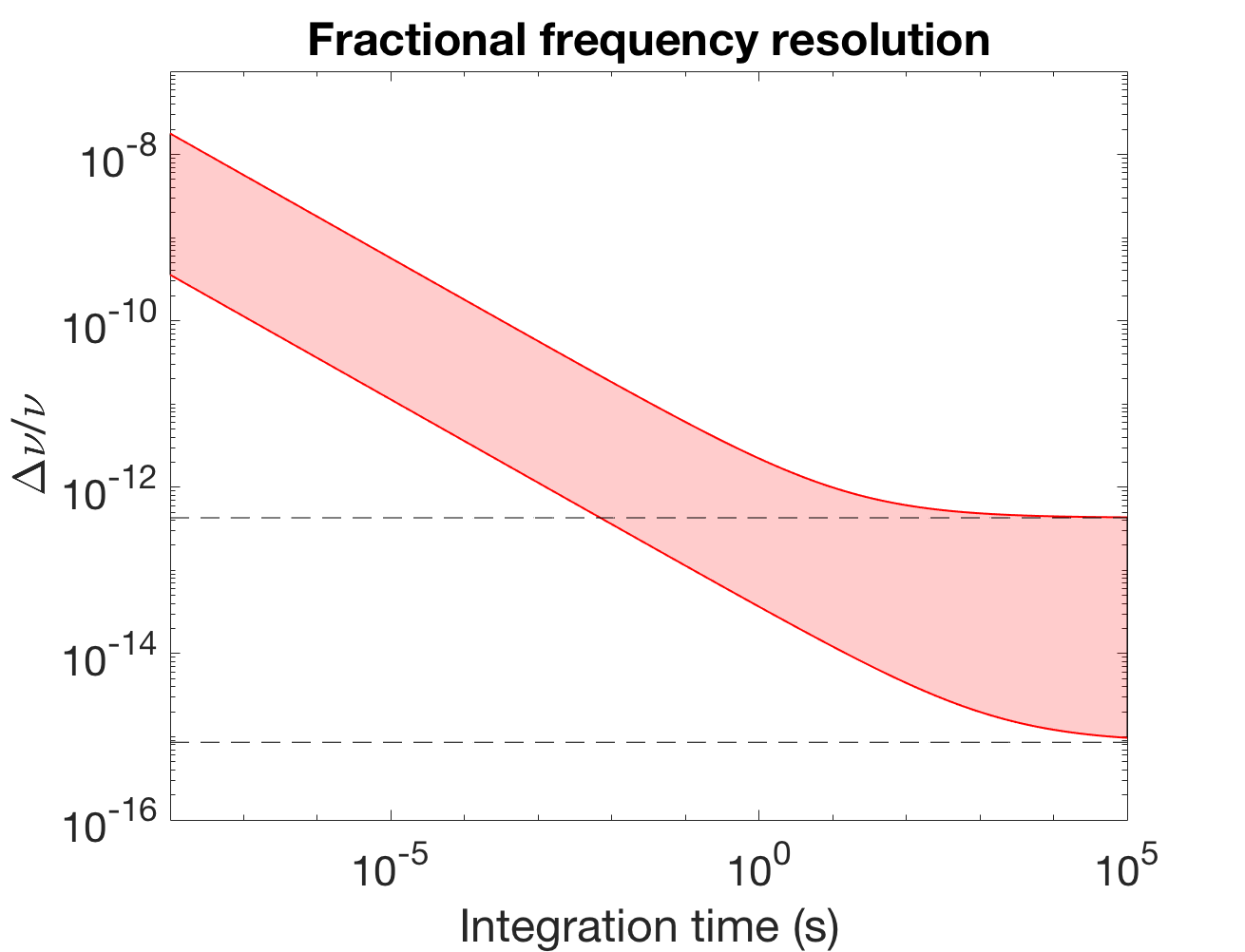}
	\caption{Fractional frequency resolution of the polaritonic clock. The top bound is given by the current parameter set in Table 1, and the lower bound by the outlook parameters. Dashed lines indicate the thermal-stability floors.}
\end{figure}

The optical spin polarization produced under off-resonant laser pumping has several implications for clock performance. Using a simple rate equation model, the steady-state spin polarization (defined as the fraction of total population in the $|0\rangle$ state) is given by $P = \frac{\gamma+g_0\alpha}{\gamma+2g\alpha+\gamma_0}$ where $\gamma_0 = \frac{1}{T_1}$ is the spin relaxation rate in the absence of the cavity microwave drive (Rabi frequency $g\alpha$) and optical polarization (rate $\gamma$). Following equation 7, the requirement of low spin excitation $P \sim 0$ results in a tradeoff between the intracavity microwave field power $\beta$ and optical polarization rate $\gamma$. This in turn results in a tradeoff between power consumption (dominated by the laser pump) and measurement SNR. Current NV sensors use laser powers on the order of watts to achieve $\beta^2 \sim10^{20}$ photons/sec. An increase in the single spin-cavity coupling $g_j$, for example via a reduced cavity mode volume, would allow for improved measurement SNR with fixed power budget.  

Second, fluctuations in laser power can change the net spin polarization, resulting in a change in coupling $g$ and thus polariton frequency. In steady-state, the change in spin polarization with optical polarization rate is $dP/d\gamma = \frac{\gamma_0}{(\gamma+g\alpha+\gamma_0)^2}$. NV ensembles have lifetimes in the 10 millisecond range under ambient conditions\cite{Jarmola2012-ht}, compared to the microsecond optical pumping rates that can be achieved without line broadening, resulting in a fractional change in coupling with optical power fluctuations $dg/g \sim 10^{-8} d\gamma/\gamma$. Fractional stability on the order of 1 ppm, which can be achieved via active feedback~\cite{Tricot2018-vm} is therefore required to reach thermal limits.

Several additional elements can influence clock stability. Magnetic fields can cause a resonance shift, although they are suppressed to first order due to the spin-1 nature of the NV. The frequency shift for a given magnetic field is shown in Figure 2d, showing this first-order cancellation, and as a function of coupling strength in Figure 3d. To reach the thermal limits described above, magnetic fluctuations should be reduced below 10 nT. This likely necessitates the use of passive shielding as in other compact atomic clocks. The diamond and cavity must be in thermal equilibrium for the polaritonic compensation to function, which requires slow thermal shifts. Aging of the microwave cavity could also result in a change in spin-cavity detuning and a polariton shift from the desired operating point. This could be detected by probing both polariton branches $\nu_\pm$ and tracking their relative frequency, followed by a re-calibration if required. Changes in strain, including vibrations, could potentially affect clock performance, although the high frequency of the clock (far above mechanical resonances), and high Young's modulus of diamond suppresses these effects. Higher-order dependencies of the NV and microwave cavity with temperature could play a role as the first-order effects are eliminated, and non-Gaussian ensemble inhomogeneity could result in a complex, potentially time-varying resonance lineshape. These effects will be important considerations in the experimental realization of the PSSC, and will bound the achievable long-term accuracy. 

The achievable shot-noise-limited SNR in this analysis is bounded by the assumption of low excitation of the spin population. Beyond this regime, however, pulsed operation of the clock becomes possible as coherent manipulation faster than dephasing or cavity loss is achievable. The use of dynamical decoupling sequences could allow for a reduction in NV linewidths towards the kHz regime, comparable to Cs clocks. Further advanced schemes based on two probe fields e.g. a coherent population trapping clock\cite{Vanier2005-tv} would also be possible in this limit, as well as schemes involving collective non-classical behavior such as ensemble squeezing\cite{Bennett2013-bt,Xia2016-hm,Zhu2014-iu}. The continued improvement in coupling strengths or decrease in cavity loss rates will yield a net improvement in performance, but will require a control scheme more advanced than the CW approach.



\begin{table}[t]
\caption{\label{tab:params}%
Sample parameters for the polaritonic clock.
}
\begin{ruledtabular}
\begin{tabular}{lccc} 
\textrm{Parameter}  & \textrm{Symbol} & \textrm{Current} & \textrm{Outlook} \\
\colrule
 Cavity output rate (kHz) & $\kappa$ & $2\pi\cdot200$  & $2\pi\cdot50$  \\ 
 NV dephasing rate (MHz) & $\Gamma$ & $2\pi\cdot3$  & $2\pi\cdot1$ \\
 Cavity-NV coupling (MHz) & $g$ & $2\pi\cdot1$  & $2\pi\cdot5$ \\
 Temperature sensitivity ratio & $R$ & -0.1 & -0.05 \\
 Single cavity-NV coupling (Hz) & $g_0$ & 0.1  & 0.3  \\
 Number of NV centers & $N$ & $2.5\cdot10^{14}$ & $4\cdot10^{14}$  \\
  Input MW power (photons/sec) & $I$ & $1\cdot10^{18}$ & $1\cdot10^{20}$  \\
 Temperature Stability (mK) & $\Delta T$ & 10 & 1  \\
 
\end{tabular}
\end{ruledtabular} 
\end{table}


The concept of polaritonic stabilization is extensible to a variety of solid-state systems. Emitters in silicon carbide have also shown room-temperature optical spin polarization~\cite{Son2020-gs,Awschalom2018-en}, potentially with lower intrinsic thermal dependence than the NV center~\cite{Kraus2014-wm}, and would extend the PSSC production to the wafer scale. Other types of bosonic modes, such as mechanical resonators~\cite{Chen2019-jm}, could take the place of the microwave cavity as similar spin-boson coupling rates have been observed in spin-mechanical systems. Finally, exotic materials such as coherent magnons~\cite{Candido2020-xe} might enable the construction of similar coupled systems with engineered thermal and magnetic response.

In conclusion, we introduced the working principle of a solid-state polaritonic microwave-spin clock. The thermal shifts of both the cavity and spin system are suppressed via controlled coupling, resulting in a stability improvement of five orders of magnitude for achievable thermal noise conditions while maintaining magnetic insensitivity. Combined with the large SNR of bulk NV sensors, this results in fractional frequency deviations that exceed current chip-scale atomic clocks and mechanical oscillators. Dispensing of the need for precision optical elements, combined with ambient operation, indicates that the PSSC platform has a path for significant advantages in size, weight and power. Following work on CMOS integration and miniaturization of the diamond platform\cite{Kim2019-bw}, polaritonic quantum clocks promise to enable a range of new applications requiring atomic-like timekeeping in a solid-state device.

\section{Acknowledgements} The authors thank Erik Eisenach for his insights regarding cavity coupled diamond ensembles pertaining to clock operation and Victor Acosta for thoughtful discussions. M.T. acknowledges support through the Army Research Laboratory ENIAC Distinguished Postdoctoral Fellowship. This work was funded in part by the Army Research Office MURI ``Ab-Initio Solid-State Quantum Materials'' (W911NF1810432), an Army Research Office DIRA ECI grant, and the NSF Center for Ultracold Atoms.

\bibliography{references}

\end{document}